\documentclass[preprintnumbers,secnumarabic,amsmath,amssymb,nofootinbib,floatfix,superscriptaddress]{revtex4}
\usepackage{graphicx}  
\usepackage{dcolumn}   
\usepackage{bm}        
\usepackage{amssymb}   
\usepackage{amsmath}
\usepackage{slashed}
\usepackage{hyperref}
\usepackage{color}
\usepackage{epsfig}
\usepackage{comment}
\usepackage{epstopdf}
\usepackage[T1]{fontenc} 
\begin{document}

\hspace{5.2in} \mbox{ACFI-T15-03}
\title{{\bf The axial anomaly, dimensional regularization and Lorentz-violating QED}}

\author{Basem Kamal El-Menoufi}
\affiliation{Department of Physics,
University of Massachusetts\\
Amherst, MA  01003, USA}
\author{G. A. White}
\affiliation{Department of Physics,
University of Massachusetts\\
Amherst, MA  01003, USA}
\affiliation{School of Physics and Astronomy,
Monash University\\
Vicotria 3800, Australia}

\begin{abstract}
In order to treat loops in the Lorentz-violating QED model, we present a derivation of the QED axial anomaly that specifically highlights the {\em infrared} origin of the effect. This is done using dimensional regularization while treating $\gamma_5$ as a spectator. This enables us to revisit aspects in the fermionic sector of Lorentz-violating QED which have analogous structure to the axial anomaly. In particular, it is shown that both the Chern-Simons and photon mass operators are not induced in the one loop effective action. At lowest order in the Lorentz-violating parameter, we can define a non-relativistic potential that captures the effects of vacuum polarization. This leads to a Zeeman-like effect in vacuum which lifts the two-fold degeneracy of the hydrogen atom S-orbitals.
\end{abstract}

\maketitle

\section{Introduction}
Since the advent of quantum mechanics, the role symmetries play in constructing fundamental theories has been paramount. Lorentz invariance is the spacetime symmetry underlying our most successful theory of nature; the Standard Model of particle physics (SM). Despite its great successes, the study of possible violations of the symmetry structure of the model is a desirable endeavor. In general, the search for symmetry violations is one possible probe of new physics beyond the SM \cite{Samuel,Potting}. In partiulcar, Lorentz invariant theories with extra dimensions must incorporate a viable mechanism by which the symmetry gets spontaneously broken at low energies \cite{Arkani1,Arkani2}. An extra motivation comes from the proposal of emergent symmetries \cite{Ambjorn,Gu,Levin,Lee,Seiberg,Sindoni,Tate}. One can envisage that the UV completion of the SM does not possess any of its observed symmetries and that the symmetry structure manifests itself only at low energies in the effective theory. In any scenario the symmetry is not expected to be exact and possible violations should be present even at low energies.

Although the specific nature of symmetry violation is important on a fundamental level, the phenomenological implications of such violations can be systematically studied within an effective field theory (EFT) framework. This approach has opened the door for the construction of the Standard Model Extension (SME) \cite{Colladay1, Colladay2}, which is an extension of the SM to include all possible Lorentz-violating renormalizable operators composed from the SM fields and respecting the SM gauge symmetries. The theory nevertheless respects the so-called observer Lorentz transformations, namely invariance under the conventional boosts and rotations of special relativity. The violation comes about by the existence of tensor fields associated with the spacetime vacuum structure. These ultimately break rotational and boost invariance in any specific fixed frame which leads observables to depend on the orientation/speed of the experimental set-up. These tensors were shown to arise naturally in string theory \cite{Samuel}. A similar approach was used in \cite{Basem1} to study possible violations of the $U(1)_{EM}$ gauge symmetry of the SM.

On the other hand, some classical symmetries may not survive after a theory is quantized. The most famous examples being the axial anomaly in the SM and the trace anomaly in conformal field theories. Although the study of anomalies remains an active area of research, by now the physics of anomalies is covered in most textbooks on quantum field theory. In perturbation theory, the traditional derivations tend to focus on the {\em ultraviolet} behavior of Feynman diagrams. However, through the use of dispersive techniques Dolgov and Zakharov \cite{Dolgov} followed by Frishman et al. \cite{Banks} emphasized the {\em infrared} origin of the axial anomaly\footnote{See also \cite{Horejsi,Schnabl,Giannotti} for dispersive derivations of anomalies that highlights the same physics.}. In the context of the trace anomaly, the same physics was elucidated in \cite{Basem2} through the explicit construction of the low-energy effective action that encodes the long-distance fluctuations of massless particles. With this physics in mind, we rederive the axial anomaly of massless QED \cite{Adler,Bell,Bardeen} using the more familiar dimensional regularization (DR). To highlight the key physics, no reference is being made to the definition of $\gamma_5$ in $D$ dimensions. The axial anomaly should not be affected by the usual subtleties associated with $\gamma_5$ since it genuinely emerges from the infrared. Our derivation is instructive and, as we explain below, proves essential to discuss the Lorentz-violating QED model.   

The derivation of the axial anomaly opens the door to re-investigate certain aspects in Lorentz-violating QED (LVQED) that share identical structure with the perturbative analysis of the axial anomaly. Those features of the model were the subject of an enormous body of work, see for instance \cite{Jackiw1,Perez1, Perez2, Andrianov, Jackiw2, Bonneau, Chung1, Chung2, Ma, Chen, Battistel, Alfaro,Altschul1,Altschul2} which is by no means a complete list. Primarily, we focus on the possible induction of the Chern-Simons and photon-mass operators in the one-loop effective action. We argue that the ability of our method to capture the infrared structure of the axial anomaly provides a thorough test for the physical significance of the previously obtained results. The consistent implementation of DR in the presence of $\gamma_5$ enables us to reliably tackle a rather peculiar feature of the Lorentz-violating QED, namely the radiative breakdown of gauge invariance which was first realized in \cite{Altschul1,Altschul2}. 

In general, the study of possible symmetry violations in electromagnetic phenomena is arguably the best way to constrain the parameter space of the SME. One typically expects that tree level processes lead to the tightest bounds, but in some cases loop effects might also become important offering in addition interesting insights about the structure of the theory. In this work, our focus is on a special operator in the fermionic sector of LVQED \cite{Colladay1}
\begin{align}\label{lagr}
\mathcal{L} = \mathcal{L}_{QED} - \bar{\psi} \slashed{b} \gamma_5 \psi
\end{align}     
where $\psi$ is a four component Dirac spinor representing any charged lepton or quark and $b^\mu$ is a spacetime-independent four-vector parameterizing Lorentz violation. The new term is $CPT$-odd since $b^\mu$ is invariant under $CPT$ \cite{Colladay2}. Moreover, the $U(1)_{EM}$ gauge symmetry of electromagnetism is manifest.
 
We will consider the one loop effective action (EA) including the vacuum polarization effects. It was first noticed in \cite{Colladay1} that the symmetry-breaking term could potentially induce the Chern-Simons (CS) electromagnetic operator in the EA\footnote{The authors of \cite{Mariz,Felipe} studied the theory over curved backgrounds.}. The CS operator reads\footnote{The classical properties of such operator are very interesting and were discussed thoroughly in the seminal work of \cite{Carroll}. Notice also that the CS operator changes by a total derivative under a gauge transformation, i.e. the Lagrangian density is not gauge-invariant.}
\begin{align}
\mathcal{L}_{CS}[A] =  \frac{1}{2} k_\mu A_\nu \tilde{F}^{\mu\nu}
\end{align}
where $k^{\mu}$ is a constant four-vector proportional to $b^\mu$. Nevertheless, the authors pointed out that the result is regularization-dependent. At $\mathcal{O}(b^\mu)$, the structure of the vacuum polarization tensor is in close analogy with the axial anomaly graphs. However, the result is manifestly gauge-invariant and thus it is not possible to fix the ambiguity in the induced coefficient by a symmetry requirement. It was later argued in \cite{Jackiw1} that a {\em non-perturbative} treatment, which we review in the body of the paper, could fix this ambiguity thus assigning a preferred value for the induced coefficient. Another striking feature of this particular treatment is the violation of the Ward-Takahashi identity which takes place at $\mathcal{O}(b^2)$ \cite{Altschul1,Alfaro}. It is needless to mention that the loss of transversality threatens the consistency of the theory and its renormalizability. We critically review the non-perturbative formulation to argue against a particular aspect of the construction.

After developing our DR-based method for the axial anomaly, we apply it to the vacuum polarization in LVQED. We find that both operators do not arise at one loop, in particular, gauge invariance is manifest. Our approach substantially differs from the previous literature in that it shifts the attention from the regularization issues to focus more on the physical content of the results. In particular, it enables us to raise and answer the question if these {\em anomalous} operators arise from the calculable low-energy physics. In addition, the consistent treatment of dimensionally regularized loops that contain $\gamma_5$ is essential to discuss the alluded to aspects of LVQED. 

Finally, the vacuum polarization contains finite pieces proportional to the Lorentz-violating parameter. At leading order in $b^\mu$ and below the electron mass, higher derivative Chern-Simons-like operators are induced in the EA. This class of corrections have not received much attention in the literature. We find that they lead to distinct phenomenology which will be discussed in the context of the hydrogen atom. In particular, the Lorentz-violating background introduces a Zeeman-like effect in vaccum which splits the hydrogen atom ground state as follows
\begin{align}
\Delta E^{(1)}_{g.s.} = \frac{8 \alpha^2 |\mathbf{b}|}{9 \pi \, a_0^3 \, m^3}
\end{align}   
where $a_0$ is the Bohr radius and $m$ is the electron mass. This correction can be used to set a bound on $\mathbf{b}$ possibly stronger than the current bounds available in the literature.

The plan of the paper is the following. In section \ref{chiral}, we re-derive the QED axial anomaly as the primary example of our method and to glean the main message of our analysis. In section \ref{disc}, we review the non-perturbative formulation of the theory and point out a common inconsistency in the original computations which precisely lead to the appearance of the CS operator and the violation of the Ward identity. We then use our method to show that both effects do not arise, in particular, gauge invariance is manifest. In section \ref{new} we use the non-vanishing part of the vacuum polarization at $\mathcal{O}(b^\mu)$ to compute the correction to the Coulomb potential which leads to a Zeeman-like effect in vacuum. Finally in section \ref{conc} we conclude and summarize our results.

\section{The axial anomaly using dimensional regularization}\label{chiral}

The original derivation of the axial anomaly \cite{Adler,Bell,Bardeen} is carried in $4$-dimensions without the explicit use of a regulator. If on the other hand an attempt is made to use DR, one naively worries about handling $\gamma_5$\footnote{For a thorough review, see \cite{Jeger,Ferrari}.}. In this section, we show that the axial anomaly is correctly reproduced assuming complete ignorance about the properties of $\gamma_5$ in $D$-dimensions. Anomalies in field theory are insensitive to the UV \cite{Basem2}, i.e. the result is unchanged by the unknown heavy physics that operates at short-distance. In particular, one should find this echoed by the insensitivity of the axial anomaly to the specific definition of $\gamma_5$ when DR is employed. In general, a regularization prescription is a deformation of the original theory in order to control the ultraviolet behavior. The exact prescription of $\gamma_5$ is part of the regularization procedure and thus one should expect to find the axial anomaly free of any dependence on the the latter. Indeed this is the case as we show next.

Let us start by reminding that the classical Lagrangian of massless QED has a global chiral symmetry
\begin{align}
\psi \rightarrow e^{i\alpha \gamma_5} \psi \ \ .
\end{align}
The Noether current associated with this symmetry reads
\begin{align}
\quad J_5^\mu = \bar{\psi} \gamma^\mu \gamma_5 \psi
\end{align}
and is conserved upon using the equations of motion. In perturbation theory, we can calculate the vacuum matrix element of the axial current to one-loop order
\begin{align}
\langle 0 | J_5^\mu(x) | 0 \rangle = \int \,\frac{d^Dp}{(2\pi)^D}\frac{d^Dk}{(2\pi)^D}\, e^{iq\cdot x} A_\nu(p) A_\lambda(k) \mathcal{M}^{\mu\nu\lambda}, \quad q=k+p \ \ .
\end{align}
The amplitude is given by two diagrams
\begin{figure*}[ht]
\centering
\includegraphics[width=0.5\textwidth]{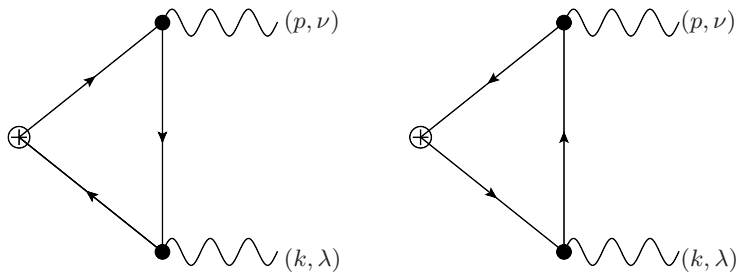}
\put(-150,90){$(p,\nu)$}
\put(-150,0){$(k,\lambda)$}
\put(0,90){$(p,\nu)$}
\put(0,0){$(k,\lambda)$}
\caption{Triangle diagrams relevant for the axial current matrix element.}
\label{tr}
\end{figure*}
and the second diagram is obtained from the first by merely interchanging $(p,\nu)$ and $(k,\lambda)$. Hence, it suffices to compute
\begin{align}
\mathcal{M}_{(1)}^{\mu\nu\lambda}(k,p) = -i e^2 &\int \frac{d^Dl}{(2\pi)^D} \bigg\{ \gamma^\mu \gamma_5 \frac{\slashed{l}-\slashed{k}}{(l-k)^2+i0} \gamma^\lambda \frac{\slashed{l}}{l^2+i0} \gamma^\nu \frac{\slashed{l}+\slashed{p}}{(l+p)^2+i0}\bigg\} \ \ .
\end{align}
where the photons are on-shell $p^2 = k^2 = 0$ and the curly brackets denote a trace operation. We integrate over the loop momentum via conventional means and contract with $q_\mu$ to find the anomalous contribution to the divergence of the axial current
\begin{align}
iq_\mu \mathcal{M}_{(1)}^{\mu\nu\lambda} = \mathcal{A}^{\nu\lambda}(k,p) + \mathcal{B}^{\nu\lambda}(k,p) 
\end{align}
where
\begin{align}\label{Atensor}
\nonumber
\mathcal{A}^{\nu\lambda}(k,p) =& \frac{ie^2}{32\pi^2} \int_0^1 dx \int_0^{1-x} dy \, \left[\frac{1}{\bar{\epsilon}} - \ln \left(\frac{\Delta}{\mu^2}\right)\right] \{\slashed{q} \gamma_5 \gamma^\alpha \gamma^\lambda \gamma_\alpha \gamma^\nu (x\slashed{k} + (1-y) \slashed{p}\\
&- \slashed{q} \gamma_5 (y\slashed{p} + (1-x)\slashed{k}) \gamma^\lambda \gamma^\alpha \gamma^\nu \gamma_\alpha
- \slashed{q} \gamma_5 \gamma^\alpha \gamma^\lambda (y\slashed{p}-x\slashed{k}) \gamma^\nu \gamma_\alpha\}
\end{align}
and
\begin{align}
\mathcal{B}^{\nu\lambda}(k,p) &= -\frac{ie^2}{16\pi^2} \int_0^1 dx \int_0^{1-x} dy \, \frac{ \{\slashed{q} \gamma_5 ((x-1)\slashed{k}-y\slashed{p})\gamma^{\lambda} (x\slashed{k}-y\slashed{p}) \gamma^\nu (x\slashed{k}+(1-y)\slashed{p})\}}{\Delta}
\end{align}
and we defined
\begin{align}
\Delta = -q^2 x y, \quad \frac{1}{\bar{\epsilon}} = \frac{1}{\epsilon} - \gamma + \ln 4\pi \ \ .
\end{align}
Notice that the tensor $\mathcal{B}^{\nu\lambda}(k,p)$ is UV finite, and naively appear to have an infrared singularity. Nevertheless, this tensor is infrared safe as we show below. We now come to a central point of our derivation, namely that all UV divergences cancel without any reference to the definition of $\gamma_5$ in $D$-dimensions. In other words, $\gamma_5$ is treated as a spectator. Let us isolate the divergent piece in the tensor $\mathcal{A}^{\nu\lambda}$ and employ the usual identities
\begin{align}\label{ddimid}
\gamma^\alpha \gamma^\mu \gamma_\alpha = (-2+2\epsilon)\gamma^\mu, \quad \gamma^\alpha \gamma^\mu \gamma^\lambda \gamma^\nu \gamma_\alpha = -2 \gamma^\nu \gamma^\lambda \gamma^\mu + 2\epsilon \gamma^\mu \gamma^\lambda \gamma^\nu
\end{align} 
along with the cyclic property of traces to find
\begin{align}\label{aeps}
\mathcal{A}_{div.}^{\nu\lambda} = \frac{ie^2}{96\pi^2 \bar{\epsilon}} \left[4 p^\lambda \{\gamma^\nu \slashed{q} \gamma_5\} - 4 k^\nu \{\gamma^\lambda \slashed{q} \gamma_5\} - 2 p^\nu \{\gamma^\lambda \slashed{q}\gamma_5\} + 2 k^\lambda \{\gamma^\nu \slashed{q} \gamma_5\} + 2 g^{\nu\lambda} \{ \slashed{k}\slashed{p} \gamma_5 - \slashed{p}\slashed{k}\gamma_5\}\right] \ \ .
\end{align} 
The above identities and trace cyclicity are true statements in $D$-dimensions. Notice first of all that traces with four Dirac matrices cancel out. Now the interesting feature is that the divergent tensor is anti-symmetric under the interchange $(p,\nu)$ and $(k,\lambda)$ and thus upon adding the two diagrams we find
\begin{align}\label{uvcancel}
\mathcal{A}_{div.}^{\nu\lambda}(k,p) + \mathcal{A}_{div.}^{\lambda\nu}(p,k) = 0 \ \ .
\end{align}
At this stage the result is proven to be {\em unambiguously} finite and regularization-independent. In general, if one defines a certain extension of $\gamma_5$ then any trace becomes an analytic function of $\epsilon$
\begin{align}
\bigg\{ \prod_{i=1}^m \gamma^{\mu_i} \gamma_5 \bigg\} = \sum_n T^{\mu_1 ... \mu_m}_{(n)} \epsilon^n
\end{align}
where the tensors appearing in the expansion will depend on the prescription used for $\gamma_5$. If the symmetry structure of the divergent tensor in eq. (\ref{aeps}) was different such that eq. (\ref{uvcancel}) is not satisfied then the final answer, despite being finite, would explicitly depend on the specific $\gamma_5$-prescription.

Now we can freely enforce the following limits on the traces 
\begin{align}\label{limit}
\lim_{\epsilon \rightarrow 0} \{\gamma^\mu \gamma^\nu \gamma^\alpha \gamma^\beta \gamma_5\} = -4i \epsilon^{\mu\nu\alpha\beta}, \quad \lim_{\epsilon \rightarrow 0} \{\gamma^\mu \gamma^\nu \gamma_5\}= 0, \quad \mu,\nu,\alpha,\beta \in 0,1,2,3 \ \ .
\end{align}
These limits must exist regardless of any $\gamma_5$-prescription used, in particular, they ensure that the result of any finite matrix-element in the theory computed with DR is identical to the 4$D$ result. We show next that indeed the anomaly is recovered using the above limits.

Turning our attention to the finite part of $\mathcal{A}^{\nu\lambda}$, we see that it comes comes from the $\mathcal{O}(\epsilon)$ pieces in eq. (\ref{ddimid}) and the $\ln \Delta$ piece in eq. (\ref{Atensor}). Hence
\begin{align}
\lim_{\epsilon \rightarrow 0} \left[\mathcal{A}^{\nu\lambda}(k,p) + \mathcal{A}^{\lambda\nu}(p,k)\right] = \frac{e^2}{6 \pi^2} \epsilon^{\lambda\nu\alpha\beta} p_\alpha k_\beta \ \ .
\end{align}
Moving to $\mathcal{B}^{\nu\lambda}(k,p)$, we find the expected feature that once the limits in eq. (\ref{limit}) are enforced the potentially singular terms vanish identically and one finds \footnote{Notice that this is indeed well-defined since the trace operation formally precedes performing the Feynman integrals.} 
\begin{align}
\lim_{\epsilon \rightarrow 0} \left[\mathcal{B}^{\nu\lambda}(k,p) + \mathcal{B}^{\lambda\nu}(p,k)\right] = \frac{e^2}{3 \pi^2} \epsilon^{\lambda\nu\alpha\beta} p_\alpha k_\beta \ \ .
\end{align}
Putting everything together yields the well-known result
\begin{align}
\langle 0 | \partial \cdot J_5 | 0 \rangle = \frac{e^2}{2\pi^2} \epsilon^{\mu\nu\alpha\beta} \int \,\frac{d^Dp}{(2\pi)^D}\frac{d^Dk}{(2\pi)^D}\, e^{iq\cdot x} A_\nu(p) A_\lambda(k) \, p_\alpha k_\beta \ \ .
\end{align}
This is one of our main results. Contrary to the naive expectation, we showed that using DR even in the presence of $\gamma_5$ is innocuous if one is interested in an {\em infrared} effect such as the axial anomaly. Our method should be applicable to any computation sharing similar structure with the axial anomaly example.

\section{The Chern-Simons term and the Ward-Takahashi identity}\label{disc}
In this section, we apply the method of the previous section to Lorentz-violating QED given by the Lagrangian in eq. (\ref{lagr}). In particular, we focus on the vacuum polarization contribution to the one loop effective action up to quadratic order in the Lorentz violating vector $b^\mu$. Our concern is two-fold; the value of the Chern-Simons coefficient and whether the Ward identity is violated. Before we do so, let us sketch the main ideas presented in the literature and point out an inconsistency in the original derivations.

The first attempt to compute the Chern-Simons coefficient appeared in \cite{Jackiw1} which was followed by plenty of discussions regarding the (un)-ambiguity of the induced operator, see for example \cite{Perez1, Perez2, Andrianov, Jackiw2, Bonneau, Chung1, Chung2, Ma, Chen, Battistel}. In \cite{Jackiw1}, the authors introduced a novel {\em non-perturbative} formulation of the theory to argue for a preferred non-vanishing value for the induced coefficient. Non-perturbative simply means that one uses the exact fermion propagator rather than treating ($-i\slashed{b}\gamma_5$) perturbatively as a new vertex in the theory. For example, the vacuum polarization (VP) tensor becomes
\begin{align}\label{VP}
i\Pi^{\mu\nu}(p) = e^2 \int\frac{d^4l}{(2\pi)^4} \, \bigg\{ \gamma^\mu \frac{i}{\slashed{l} - m - \slashed{b}\gamma_5} \gamma^\nu \frac{i}{\slashed{l} + \slashed{p} - m - \slashed{b}\gamma_5} \bigg\} \ \ .
\end{align}

In the conventional perturbative treatment the Chern-Simons coefficient is finite but ambiguous \cite{Colladay1,Jackiw2}. To linear order in $b^\mu$, there are two diagrams shown in figure 
with each being linearly divergent but the final result is finite. Nevertheless, one is free to shift the momentum variable in each diagram by different amounts and, due to the linear divergence, this renders the final answer undetermined. The issue here is very similar to the axial anomaly graphs, but with one subtle distinction: the answer, being proportional to the Levi-Civita tensor, is gauge invariant. In conclusion, there is no symmetry to impose on the matrix element which leaves the induced coefficient ambiguous.
\begin{figure*}[ht]
\centering
\includegraphics[width=0.5\textwidth]{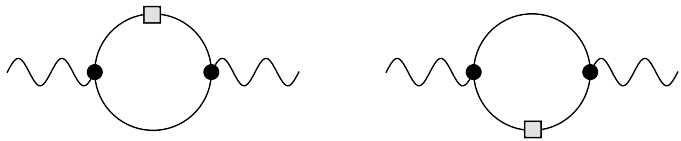}
\caption{The $\mathcal{O}(b)$ contribution to the vacuum polarization. The gray square denotes the vertex $-i\slashed{b}\gamma_5$.}
\label{pbubbles}
\end{figure*} 

On the contrary, there is a single diagram in the non-perturbative treatment given by eq. (\ref{VP}). Practically, one expands the propagators in powers of $b^\mu$ to perform the computation and evidently the same integrals at $\mathcal{O}(b)$ arises similar to the perturbative treatment. The new feature is that the momentum routing between the two integrals becomes correlated \cite{Jackiw1,Jackiw2} which gives a preferred value for the induced coefficient. Subsequently and using different regulators, a variety of answers were displayed in the literature \cite{Perez1, Perez2, Andrianov, Jackiw2, Bonneau, Chung1, Chung2, Ma, Chen, Battistel}. A shocking aspect of this formulation is the violation of the Ward identity at $\mathcal{O}(b^2)$ \cite{Altschul1,Altschul2,Altschul3}.

Our primary goal in this section is to highlight an inconsistency in the original computations, in particular, the misuse of {\em symmetric integration} in divergent integrals. To keep our presentation self-contained, we review the calculations presented in \cite{Jackiw1,Altschul1} in appendix \ref{nonper}. These results hinged on the use of symmetric integration by which logarithmic divergences cancel out and one can evaluate all integrals in 4$D$ without an explicit use of a regulator. By examining a well-known integral, we show below that {\em symmetric integration} is not a valid operation in the presence of divergences. We then move to apply our method to these calculations.

\subsection{The problem with symmetric integration}

Let us consider the following integral in 4$D$
\begin{align}\label{integral}
I_\mu(p) = \int \, d^4l \left[\frac{l_\mu}{l^4} - \frac{(l-p)_\mu}{(l-p)^4}\right] \ \ .
\end{align}
This integral is linearly divergent and appears to vanish under a naive shift of the integration variable. Nevertheless, a careful treatment yields an unambiguous finite answer. We start by defining
\begin{align}
p_0 = i p_4, \quad l_0 = i l_4, \quad I_0 = iI_4 \ \ .
\end{align}
We then Wick rotate to go to Euclidean space. To turn the difference in the integrand into a surface term, the second term is Taylor expanded
\begin{align}\label{expansion}
\frac{(l-p)_i}{(l-p)^4} = \frac{l_i}{l^4} - p^j \partial_j \frac{l_i}{l^4} + \frac{1}{2} p^j p^k \partial_j \partial_k \frac{l_i}{l^4} + ... \ \ .
\end{align}
By Gauss's theorem, the 4$D$ integral is turned into a surface integral over a 3-sphere at infinity and thus all terms with two or more derivatives vanish. We find
\begin{align}
I_i (p) = i p_j \lim_{l\rightarrow \infty}\oint d\Omega_3 \frac{l_i l_j}{l^5} \ \ .
\end{align}
Spherical symmetry allows us to replace $l_i l_j = \delta_{ij}l^2/4$ and the integral becomes elementary. Transforming back to Lorentzian space yields the well-known result \cite{Donoghue}
\begin{align}\label{surface}
I_\mu(p) = \frac{i \pi^2}{2}p_\mu \ \ .
\end{align}

On the other hand, one can arrange for this integral to vanish if symmetric integration is used instead. To see how this comes about, let us substitute the expansion of eq. (\ref{expansion}) in eq. (\ref{integral}) and discard the quadratic pieces that do not contribute. Instead of using Gauss's theorem, one can identically write
\begin{align}
I_i = i p^j \int \, d^4l_E \left[\frac{\delta_{ij}}{l_E^4} - \frac{4 l_i l_ j}{l_E^6}\right] \ \ .
\end{align}
Both pieces in the integrand are logarithmically divergent. If one allows symmetric integration via $l_i l_j \to l_E^2/4$ then clearly the result vanishes, in contradiction to the finite result in eq. (\ref{surface}). Since the latter is undoubted, we conclude that symmetric integration is not a well-defined operation for divergent integrals. In other words, a symmetric integration prescription in a divergent integral amounts to setting finite surface terms to zero.

\subsection{A vanishing CS coefficient and restoration of tranversality}\label{CS}

Based on the previous discussion, an explicit regulator must be introduced to carry out the computation. As we showed in the previous section, the use of DR is well suited for such computations. If the anomalous operators under study are genuinely induced by long-distance fluctuations, our method will indeed capture these effects. We will find that both operators vanish at one loop, in particular, gauge invariance is restored.

Ideally, one might hope to perform the integral in eq. (\ref{VP}) exactly and then expand the final answer in powers of $b^\mu$. The complicated form of the propagators renders this task cumbersome, although an attempt was made in \cite{Perez1}. We choose to expand the propagators directly \cite{Altschul1}
\begin{align}
\nonumber
\frac{i}{\slashed{l} - m - \slashed{b}\gamma_5} = &\frac{i}{\slashed{l} - m} + \frac{i}{\slashed{l} - m} \, (-i\slashed{b}\gamma_5) \, \frac{i}{\slashed{l} - m}\\
& + \frac{i}{\slashed{l} - m} \, (-i \slashed{b}\gamma_5) \, \frac{i}{\slashed{l} - m} \, (-i \slashed{b}\gamma_5) \, \frac{i}{\slashed{l} - m} + ... \ \ .
\end{align}  
Evidently, the use of DR ensures the result matches onto the exact computation in the appropriate limit. Thus one finds
\begin{align}
i\Pi_b^{\mu\nu}(p) = e^2 \left( \Sigma^{\mu\nu}(p) + \Sigma^{\nu\mu}(-p) \right) 
\end{align}
where
\begin{align}
\Sigma^{\mu\nu}(p) = \int\frac{d^Dl}{(2\pi)^D} \, \bigg\{ \gamma^\mu \frac{i}{\slashed{l} - m} \gamma^\nu \frac{i}{\slashed{l} + \slashed{p} - m}\, (-i\slashed{b}\gamma_5) \, \frac{i}{\slashed{l} + \slashed{p} - m} \bigg\}  \ \ .
\end{align}

Notice that a shift in one of the integrands has been used to cast the result in this form, which is indeed allowed by virtue of DR. 
It is convenient to separate $\Sigma^{\mu\nu}$ into a potentially divergent piece and a finite one
\begin{align}\label{sigma}
\Sigma^{\mu\nu} = -\frac{i}{16\pi^2} (D^{\mu\nu} + F^{\mu\nu}) \ \ .
\end{align}
Let us analyze the potentially divergent piece, it reads
\begin{align}
\nonumber
D^{\mu\nu} = \int_0^1 dx\, \left(\frac{1}{\bar{\epsilon}} - \ln \Delta \right) &\big[(-1+\epsilon)(1-x)\{\gamma^\mu \gamma^\nu \slashed{b} \gamma_5 \slashed{p} \} + (-1+\epsilon) (1-x) \{\gamma^\mu \gamma^\nu \slashed{p} \slashed{b} \gamma_5 \} \\
&+ x \{\gamma^\nu \slashed{p} \gamma^\mu \slashed{b}\gamma_5 \} - \epsilon\, x \{\gamma^\mu \slashed{p} \gamma^\nu  \slashed{b} \gamma_5 \} \big]
\end{align}
where use has been made of eq. (\ref{ddimid}) and $\Delta = m^2 - p^2 x(1-x)$. Using only the cyclic property of traces and the Dirac algebra, we find that terms multiplying $1/\epsilon$ can be reduced to
\begin{align}
D^{\mu\nu}_{div.} = \frac{1}{\bar{\epsilon}} \left[\frac{1}{3}p^\mu \{\gamma^\nu \slashed{b}\gamma_5\} + \frac{1}{3} p^\nu \{\gamma^\mu \slashed{b}\gamma_5 \} - \frac{2}{3} \eta^{\mu\nu} \{\slashed{p}\slashed{b} \gamma_5 \} \right] \ \ .
\end{align}
In particular, all traces with four Dirac matrices cancel out. The above tensor is antisymmetric under interchanging $\mu$ and $\nu$ while sending $p \to -p$, and hence $\Pi^{\mu\nu}_b$ is indeed finite. The development here is quite analogous to the axial anomaly. The finite parts are rather straightforward
\begin{align}
D^{\mu\nu}_{fin.} &= -4i \epsilon^{\mu\nu\alpha\beta} p_{\alpha} b_\beta \int_0^1 dx \, \left( x (2-3x) \ln \Delta +x^2(2x-x^2-1) \frac{p^2}{\Delta} + \frac{2}{3} \right)\\
F^{\mu\nu} &= + 4i \epsilon^{\mu\nu\alpha\beta} p_{\alpha} b_\beta \int_0^1 dx \, x(2-x) \frac{m^2}{\Delta}
\end{align}
where the limits has been taken according to eq. (\ref{limit}). To arrive at the above result, one has to be careful with terms linear in $m$. These terms comprise traces of an odd number of Dirac matrices with $\gamma_5$ but since the integrals are logarithmically divergent, their traces can not be immediately dropped. Nevertheless, a careful computation shows once again that all divergences cancel identically and hence the whole contribution vanishes when the trace limits are enforced. The Chern-Simons operator is linear in $p$, and so we take the limit $p^2 = 0$ to find
\begin{align}
\lim_{p^2 \to 0} (D^{\mu\nu}_{fin.} + F^{\mu\nu} ) = 0 \ \ .
\end{align}

There is no induced Chern-Simons term at one loop\footnote{This null result has been obained in \cite{Bonneau}.}. Notice also that DR guarantees the result is continuous in the massless limit as one can easily check by setting $m=0$ at the start\footnote{This does not take place for instance in the non-perturbative treatment as discussed in the appendix.}. We notice here that the result does not suffer from any ambiguities related to regularization and is fully consistent with gauge invariance. 

Now we move to the next term in the expansion; the piece quadratic in $b^{\mu}$ which is the source of the violation of the Ward identity. The breakdown of the Ward identity posits a serious challenge to the consistency of the theory and its renormalizability. We set $p=0$ from the start since the momentum-dependent terms necessarily satisfy the Ward identity \cite{Altschul2}. The polarization operator in eq. (\ref{VP}) expanded to second order in $b$ reads
\begin{equation}
\Pi ^{\mu \nu }_{b^2}(0) = ie^2 \left( \Omega ^{\mu \nu } + \tilde{\Omega} ^{\mu \nu} + \tilde{\Omega} ^{\nu \mu} \right)
\end{equation}
where
\begin{align}
\nonumber
\Omega^{\mu \nu } &= - \int \frac{d^Dk}{(2 \pi )^D} \frac{\{ \gamma ^\mu (\slashed{k} +m) (-i\slashed{b} \gamma _5) (\slashed{k}+m)\gamma ^\nu (\slashed{k} +m)(-i\slashed{b}\gamma _5)(\slashed{k}+m) \} }{(k^2-m^2)^4} \\
\tilde{\Omega} ^{\mu \nu} &=  - \int \frac{d^Dk}{(2 \pi )^D} \frac{\{ \gamma ^\mu (\slashed{k}+m)\gamma ^\nu (\slashed{k}+m)(-i\slashed{b} \gamma _5)(\slashed{k}+m)(-i\slashed{b} \gamma _5)(\slashed{k}+m)  \} }{(k^2-m^2)^4} 
\end{align}
We decompose the various tensors into divergent and finite pieces
\begin{eqnarray}
\Omega ^{\mu \nu } = D^{\mu \nu } + F^{\mu \nu}, \quad \tilde{\Omega} ^{\mu \nu } = \tilde{D}^{\mu \nu } + \tilde{F}^{\mu \nu},
\end{eqnarray}
with
\begin{eqnarray}
\nonumber
D^{\mu \nu } &=& \{ \gamma^\mu \gamma^\alpha (-i\slashed{b} \gamma_5) \gamma^\beta \gamma^\nu \gamma^\sigma (-i\slashed{b} \gamma_5)\gamma^\rho \} J_{\alpha \beta \sigma \rho }\\\nonumber
\tilde{D}^{\mu \nu } &=& \{ \gamma^\mu \gamma^\alpha \gamma^\nu  \gamma^\beta (-i\slashed{b} \gamma_5) \gamma^\sigma (-i\slashed{b} \gamma_5)\gamma^\rho \} J_{\alpha \beta \sigma \rho}\\\nonumber
F^{\mu \nu} &=& m^2 \{ \gamma^\mu \gamma^\alpha (-i\slashed{b} \gamma_5) \gamma^\beta \gamma^\nu (-i\slashed{b} \gamma_5) + \gamma^\mu \gamma^\alpha (-i\slashed{b} \gamma_5) \gamma^\nu \gamma^\beta (-i\slashed{b} \gamma_5) \\\nonumber
&+& \gamma^\mu \gamma^\alpha (-i\slashed{b} \gamma_5) \gamma^\nu (-i\slashed{b} \gamma_5) \gamma^\beta + \mu \leftrightarrow \nu \} J_{\alpha \beta} + m^4 \{ \gamma^\mu \gamma^\nu (- i\slashed{b} \gamma_5) (-i \slashed{b} \gamma_5) \} J_0 \\\nonumber
\tilde{F}^{\mu \nu} &=&  m^2 \{ \gamma^\mu \gamma^\alpha \gamma^\nu (-i\slashed{b} \gamma_5)\gamma^\beta  (-i\slashed{b} \gamma_5) + \gamma^\mu \gamma^\alpha \gamma^\nu (-i\slashed{b} \gamma_5)  \gamma^\beta (-i\slashed{b} \gamma_5)\\
&+& \gamma^\mu \gamma^\alpha \gamma^\nu (-i\slashed{b} \gamma_5)   (-i\slashed{b} \gamma_5)\gamma^\beta + \mu \leftrightarrow \nu \} J_{\alpha \beta }+ m^4 \{ \gamma^\mu \gamma^\nu (-i \slashed{b} \gamma_5) (-i \slashed{b} \gamma_55)  \} J_0 
\end{eqnarray}
and we defined
\begin{align}
\nonumber
J_{\alpha \beta \sigma \rho} &= \frac{-i}{24(4 \pi )^2} \left(\eta_{\alpha \beta} \eta_{\rho \sigma} + \eta_{\alpha \sigma} \eta_{\beta \rho} +\eta_{\beta \sigma} \eta_{\alpha \rho} \right) \left(\frac{1}{\bar{\epsilon}}-\log m^2 \right), \\
& J_{\alpha \beta } = \frac{i}{12(4 \pi)^2} \eta_{\alpha \beta} \frac{1}{m^2}, \quad  J_0 = \frac{i}{6 (4 \pi ) ^2} \frac{1}{m^4}
\end{align}
Notice here that we do not use the $4D$ property $\gamma_5^2 = 1$. Also, we can ignore all terms with odd powers of $m$ since they involve a trace of $\gamma_5$ and an odd number of Dirac matrices. Focusing solely on the finite tensors, we easily reproduce the result found in \cite{Altschul1}
\begin{align} \label{finiteb2}
ie^2 \lim_{\epsilon \to 0}  \left( F^{\mu \nu} + \tilde{F} ^{\mu \nu } + \tilde{F} ^{\nu \mu } \right) =  - \frac{e^2}{24 \pi ^2}\left(2 b^\mu b^\nu +\eta ^{\mu \nu} b^2 \right) \ \ .
\end{align}
We can then use the identities in eq. (\ref{ddimid}) and manipulate the traces in the divergent tensors employing the cyclic property to find that the terms proportional to $1/\epsilon$ cancel identically similar to the previous calculations. The pieces proportional to $\epsilon$ then give a finite contribution which reads
\begin{align} 
ie^2 \lim_{\epsilon \to 0} \left( D^{\mu \nu} + \tilde{D} ^{\mu \nu } + \tilde{D} ^{\nu \mu } \right)  =  + \frac{e^2}{24 \pi ^2}\left(2 b^\mu b^\nu +\eta ^{\mu \nu} b^2 \right)
\end{align}
exactly canceling the piece in eq. (\ref{finiteb2}). It is gratifying to see that gauge invariance is unambiguously manifest with our method.


\section{A Chern-Simons-type term and the hydrogen atom }\label{new}
In this section we study the contribution to $\Pi^{\mu\nu}_b$ which arises if $p^2 \neq 0$. Taking $\Pi^{\mu\nu}_b$ into account, we study the effect of the vacuum polarization on the spectrum of the hydrogen atom. The one loop vacuum polarization up to linear order in $b^\mu$ reads
\begin{align}
i\Pi^{\mu\nu} = i\Pi^{\mu\nu}_{QED} + i \Pi^{\mu\nu}_b
\end{align}
where 
\begin{align}
i\Pi^{\mu\nu}_{QED} = \frac{ie^2}{2\pi^2} \left(p^2 \eta^{\mu\nu} - p^\mu p^\nu \right) \int_0^1 dx \, x(1-x) \ln \frac{\Delta}{m^2}
\end{align}
is the usual QED vacuum polarization tensor renormalized in an on-shell scheme. We start by computing the matrix element describing electron-proton scattering assuming the proton is a Dirac fermion. The amplitude reads
\begin{figure*}[ht]
\centering
\includegraphics[width=0.3\textwidth]{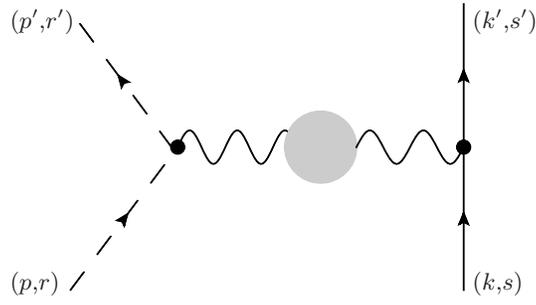}
\put(-175,1){($p$,$r$)}
\put(-175,100){($p^\prime$,$r^\prime$)}
\put(0,1){($k$,$s$)}
\put(0,100){($k^\prime$,$s^\prime$)}
\caption{Electron-proton scattering. The grey blob denotes the vaccum polarization insertion.}
\label{epscat}
\end{figure*}
\begin{align}
i\mathcal{M}_{eP} = e^2 \left[\bar{u}(k^\prime,s^\prime) \gamma^\mu u(k,s)\right] (-iG_{\mu\nu}) \left[\bar{u}(p^\prime,r^\prime) \gamma^\nu u(p,r)\right] \ \ .
\end{align}  
The different momenta and spins are labeled in figure 
Also, $G_{\mu\nu}$ is the {\em dressed} photon propagator 
\begin{align}
-iG_{\mu\nu} = \frac{-i\eta_{\mu\nu}}{q^2} + \frac{-i\eta_{\mu\alpha}}{q^2} i\Pi^{\alpha\beta} \frac{-i\eta_\beta^\nu}{q^2} + .... \ \ .
\end{align}
The dots indicate the iteration of the vacuum polarization tensor and $q = p - p^\prime = k^\prime - k$ is the momentum transfer. The appearance of the Levi-Civita tensor renders the evaluation of the above summation cumbersome and so we choose to work at fixed order and truncate the series at one loop. The one-photon exchange gives rise to the famous fine structure correction 
\begin{align}
E^{(1)}_{f.s.} = \frac{E_n^2}{2m} \left(3 - \frac{4n}{j+1/2} \right)
\end{align} 
where $j$ is the total angular momentum quantum number. Now moving to the corrections due to vacuum polarization, we find the spin-independent Uhleng potential which reads 
\begin{align}
V_{Uhleng}(r) = - \frac{e^4}{60 \pi^2 m^2} \delta^{(3)}(\mathbf{r})
\end{align} 
This potential is computed by expanding the vacuum polarization tensor in powers of $q^2/m^2$ and retaining the lowest order piece which is a good approximation since the hydrogen atom wave functions are almost constant over the electron Compton wavelength \cite{Peskin}. We do not display the shift in the spectrum due to the Uhleng term as it is irrelevant to our discussion.

What we are after is the similar corrections that arise from $\Pi_b^{\mu\nu}$. We work with spinors normalized as $\bar{u}(p,s) u(p,s) = 1$ and take the proton mass to be infinite. The photon propagator receives the following correction
\begin{align}
-i G^b_{\mu\nu} = -\frac{i}{q^2} \left(\frac{-i}{q^2} \epsilon_{\mu\nu\alpha\beta} q^\alpha b^\beta \Pi_b(q^2) \right)
\end{align}
where
\begin{align}
\Pi_b(q^2) = \frac{e^2}{2\pi^2} \int_0^1 dx \, \left[ (3x^2 - 2x) \ln \Delta + x^2 (x^2 - 2x +1) \frac{q^2}{\Delta} + x(2-x) \frac{m^2}{\Delta} - \frac{2}{3} \right] \ \ .
\end{align}   
At low momentum transfer, we expand the above function
\begin{align}
\Pi_b(q^2) \approx \frac{e^2}{12 \pi^2} \frac{q^2}{m^2} \ \ .
\end{align}
To streamline the discussion, we take the background vector to be space like, $b^\mu = (0,\mathbf{b})$. Hence, the amplitude simplifies to
\begin{align}
\mathcal{M}^b_{eP} = \frac{i e^4}{12 \pi^2 m^2} \frac{1}{\mathbf{q}^2}  \left[\bar{u}(k^\prime,s^\prime) \gamma^\mu u(k,s)\right] \epsilon_{\mu\nu ij} q^i b^j \left[\bar{u}(p^\prime,r^\prime) \gamma^\nu u(p,r)\right]
\end{align}
where the momentum transfer is purely spatial in the limit of infinite proton mass. Using some properties of the Pauli matrices, we end up with
\begin{align}
\mathcal{M}^b_{eP} = \mathcal{M}_{\slashed{s}} + \mathcal{M}_{s}
\end{align}
where
\begin{align}
\mathcal{M}_{s} = \frac{e^4}{12 \pi^2 m^2} \frac{1}{\mathbf{q}^2} \xi^{\dagger \prime}\, \frac{\sigma\cdot \mathbf{q}\, \mathbf{q} \cdot \mathbf{b} - \mathbf{q}^2 \sigma \cdot \mathbf{b}}{2m} \, \xi \, \delta^{s s^\prime}
\end{align}
and the spin-independent piece reads
\begin{align} 
\mathcal{M}_{\slashed{s}} = \frac{ie^4}{6 \pi^2 m^2} \frac{1}{\mathbf{q}^2}  \frac{ (\mathbf{b} \times \mathbf{p}) \cdot \mathbf{q}}{2m} \delta^{s s^\prime} \delta^{r r^\prime} \ \ .
\end{align}
In the Born approximation, the potential is given by the inverse Fourier transform of the matrix element
\begin{align}
V(r) = - \int \frac{d^3q}{(2\pi)^3} e^{i \mathbf{q} \cdot \mathbf{r}} \mathcal{M}(\mathbf{q}) \ \ .
\end{align}
This leads to the following potential
\begin{align}
V_{\slashed{s}}(r) &= \frac{e^4}{48 \pi^3 m^3 r^3}\, \mathbf{b} \cdot (\mathbf{p} \times \mathbf{r})\label{nospinpot} \\ 
V_s(r) &= \frac{e^4}{24 \pi^2 m^3} \left[\frac{3}{4\pi r^5} \mathbf{\sigma} \cdot \mathbf{r} \, \mathbf{b} \cdot \mathbf{r} - \frac{1}{4\pi r^3} \mathbf{b} \cdot \mathbf{\sigma} + \frac{2}{3} \mathbf{\sigma} \cdot \mathbf{b} \, \delta^{(3)}(\mathbf{r}) \right] \ \ \label{spinpot} .
\end{align}

Using lowest order perturbation theory, it is easy to obtain the shift in the spectrum of the hydrogen atom. It is ideal to only consider the correction to the S-orbitals since the spin-independent piece in eq. (\ref{nospinpot}) vanishes in this case\footnote{Notice that the expectation value of $(1/r^3)$ diverges for $l=0$ states, nevertheless, it makes sense to ignore the effect of this term based on symmetry.}. In addition, the extra spin-dependent pieces due to QED corrections are suppressed by the proton mass \cite{Donoghue}. The unique effect of the correction in eq. (\ref{spinpot}) is to lift the two-fold degeneracy of any S-wave, for example, the energy split in the ground state reads\footnote{We take the axis of quantization to be along $\mathbf{b}$ for simplicity.}
\begin{align}\label{ecorr}
\Delta E^{(1)}_{g.s.} = \frac{8 \alpha^2 |\mathbf{b}|}{9 \pi \, a_0^3 \, m^3}
\end{align}   
where $a_0$ is the Bohr radius. Hence, the Lorentz-violating correction mimics the Zeeman effect. Torsion pendulum experiments \cite{Heckel1,Heckel2} set the following bound on the Lorentz-violating parameter\footnote{See also \cite{Roberts} for bounds using atomic parity violation experiments.}: $|\mathbf{b}| \leq \mathcal{O}(1) \times 10^{-22} ev$. Although the energy splitting in eq. (\ref{ecorr}) could be used to set a direct and independent bound on $|\mathbf{b}|$, we are unfortunately not aware of any existing experiment that can proble such an effect even indirectly. This will be the subject of future work.

\section{Conclusion}\label{conc}

We used dimensional regularization to discuss a simple rather instructive derivation of the QED axial anomaly. In particular, we showed that the usual difficulties associated with $\gamma_5$ do not matter in this case. The underlying physics being the infrared origin of anomalies in field theory. We carried over the intuition to re-investigate aspects of LVQED which caused controversy in the literature. We found that the Chern-Simons coefficient unambiguously vanishes and that the Ward-Takahashi identity is preserved. The latter is very important to ensure the theory is unitary and renormalizable. We argued that our method offers a reliable test to the results previously obtained in the literature.

On the phenomenological side, we showed that the Lorentz-violating effects on the vacuum polarization give rise to Zeeman-like effect where the vector $b^\mu$ plays the role of a uniform magnetic field. Unlike the usual QED corrections, this lifts the degeneracy of the S-orbitals in the hydrogen atom. This unique feature should provide a window to set stringent limits on the Lorentz-violating parameter, a task we found tedious due to the lack of experimental input. Nevertheless, we would pursure this particular aspect in the future. Another interesting direction is to explore the predictions of the theory on gravitational phenomena especially in regard to the early Universe.   
 
\acknowledgments

We have benefited from many helpful discussions with John Donoghue and Csaba Balazs. We also thank Sean Carroll for discussions. BKE work has been supported in part by the U.S. National Science Foundation Grant No. PHY-1205896. GW would like to acknowledge J L William Scholarship, APA Scholarship and the Keith Murdoch Scholarship via the American Australian Association.
 
\appendix

\section{Non-perturbative formulation}\label{nonper}

In this appendix, we give a quick derivation of the results obtained in \cite{Jackiw1,Altschul1}. In particular, we highlight the role of symmetric integration in obtaining finite results. As previously mentioned, an exact computation (to all orders in $b^\mu$) was attempted in \cite{Perez1}. Since we are only interested in the consistency of symmetric integration, we will present the easier computation along the lines of \cite{Jackiw1}. For the convenience of the reader we adopt a similar notation despite taking a slightly different approach.
\begin{align}
i\Pi^{\mu\nu}_b = b_\alpha \Pi^{\mu\nu\alpha}(p)
\end{align}
where 
\begin{align}
\Pi^{\mu\nu\alpha}(p) = I^{\mu\nu\alpha}(p) + \tilde{I}^{\mu\nu\alpha}(p) \ \ .
\end{align}
Explicitly we have,
\begin{align}
\nonumber
I^{\mu\nu\alpha}(p) &= - \int \, \frac{d^4l}{(2\pi)^4} \, \frac{ \{ \gamma^\mu (\slashed{l} + m) \gamma^\nu (\slashed{l} + \slashed{p} + m ) \gamma^\alpha \gamma_5 (\slashed{l} + \slashed{p} + m )\}}{(l^2-m^2+i0)((l+p)^2 - m^2 + i0)^2} \\
\tilde{I}^{\mu\nu\alpha}(p) &= - \int \, \frac{d^4l}{(2\pi)^4} \, \frac{ \{ \gamma^\mu (\slashed{l} + m) \gamma^\nu (\slashed{l} + m ) \gamma^\alpha \gamma_5 (\slashed{l} + \slashed{p} + m )\}}{(l^2-m^2+i0)^2((l+p)^2 - m^2 + i0)} \ \ .
\end{align}
Recall that the above integrals do not represent two different diagrams: only global momentum shifts are allowed in the non-perturbative formulation. The remarkable feature here is that linear divergences cancel out at the integrand level as we show next. To simplify, we set $p^2 = 0$ from the start and evaluate the traces to find
\begin{align}
I^{\mu\nu\alpha}(p) = I_{\text{div.}}^{\mu\nu\alpha} + I_{\text{fin.}}^{\mu\nu\alpha}
\end{align}
where the divergent tensor reads
\begin{align}
I_{\text{div.}}^{\mu\nu\alpha}  = -4i \int \, \frac{d^4l}{(2\pi)^4} \, \frac{(l^2 + 2 l \cdot p)\epsilon^{\mu\sigma\nu\alpha} l_\sigma + 2 \epsilon^{\beta\mu\sigma\nu}l^\alpha l_\sigma p_\beta} {(l^2-m^2+i0)((l+p)^2 - m^2 + i0)^2}
\end{align}
and
\begin{align}
I_{\text{fin.}}^{\mu\nu\alpha} = \frac{1}{6\pi^2}\, \epsilon^{\mu\beta\nu\alpha} p_\beta \ \ .
\end{align}
We want to isolate the linear divergence prior to performing any loop shifts. We start by manipulating the divergent tensor 
\begin{align}\label{I}
\nonumber
I_{\text{div.}}^{\mu\nu\alpha}  = &-4i \int \, \frac{d^4l}{(2\pi)^4} \, \frac{2 \epsilon^{\beta\mu\sigma\nu}  l^\alpha l_\sigma p_\beta + m^2 \epsilon^{\mu\sigma\nu\alpha}l_\sigma} {(l^2-m^2+i0)((l+p)^2 - m^2 + i0)^2} \\
&- 4i \int\, \frac{d^4l}{(2\pi)^4} \, \frac{\epsilon^{\mu\sigma\nu\alpha}l_\sigma}{(l^2-m^2+i0)((l+p)^2 - m^2 + i0)} \ \ .
\end{align}
Similarly,
\begin{align}\label{Itilde}
\nonumber
\tilde{I}_{\text{div.}}^{\mu\nu\alpha}  = &-4i \int \, \frac{d^4l}{(2\pi)^4} \, \frac{2 \epsilon^{\beta\nu\sigma\mu}  l^\alpha l_\beta p_\sigma + l^2 \epsilon^{\nu\beta\mu\alpha} p_\beta + m^2 \epsilon^{\nu\sigma\mu\alpha}l_\sigma} {(l^2-m^2+i0)^2((l+p)^2 - m^2 + i0)} \\
&- 4i \int\, \frac{d^4l}{(2\pi)^4} \, \frac{\epsilon^{\nu\sigma\mu\alpha}l_\sigma}{(l^2-m^2+i0)((l+p)^2 - m^2 + i0)} 
\end{align}
and
\begin{align}
\tilde{I}_{\text{fin.}}^{\mu\nu\alpha} = \frac{1}{6\pi^2}\, \epsilon^{\mu\beta\nu\alpha} p_\beta \ \ .
\end{align}
Inspection of eqs. (\ref{I}) and (\ref{Itilde}) shows that the linear divergences cancel out at the integrand level and the remaining divergences are at most logarithmic. The computation proceeds by introducing Feynman parameters and performing the loop integral. So far we have not introduced any kind of regulators, and here is where symmetric integration kicks in. Using the latter prescription, one readily finds that the logarithmic divergences cancel out at the integrand level as well. After this cancellation, we can collect the finite pieces to find
\begin{align}
\Pi^{\mu\nu\alpha}(p) = \frac{3}{8\pi^2} \, \epsilon^{\mu\beta\nu\alpha}p_\beta
\end{align}
identical to the result in \cite{Jackiw1}. Notice that if the fermion mass was set to zero at the start, the result becomes
\begin{align}
\Pi^{\mu\nu\alpha}(p;m=0) = - \frac{1}{8\pi^2} \, \epsilon^{\mu\beta\nu\alpha}p_\beta \ \ .
\end{align}
This discontinuity alone is enough to suspect the validity of the procedure. A similar story takes place for $\Pi^{\mu\nu}_{b^2}(0)$ \cite{Altschul1} with the end result displayed in eq. (\ref{finiteb2}). We do not include the full derivation since it is very similar to the above computation.

\end{document}